\documentclass[a4paper]{jpconf}
\usepackage[english]{babel}
\usepackage{graphicx}
\usepackage{psfrag}
\begin{document}
\title{Production of a forward $J/\psi$ and a backward jet at the LHC}

\author{R Boussarie$^1$, B Duclou\'{e} $^{2,3}$, 
L Szymanowski$^{1,4,5}$ and S Wallon$^{1,6}$}

\address{$^1$ Laboratoire de Physique Th\'{e}orique, CNRS, Universit\'{e} Paris Sud, Universit\'{e} Paris Saclay, 91405 Orsay, France}
\address{$^2$ Department of Physics, University of Jyv\"{a}skyl\"{a}, P.O. Box 35, 40014 University of Jyv\"{a}skyl\"{a}, Finland}
\address{$^3$ Helsinki Institute of Physics, P.O. Box 64, 00014 University of Helsinki, Finland}
\address{$^4$ National Centre for Nuclear Research (NCBJ), Warsaw, Poland}
\address{$^5$ Centre de Physique Th\'eorique, Ecole Polytechnique, CNRS, Universit\'{e} Paris Saclay, F91128 Palaiseau, France}
\address{$^6$ UPMC Universit\'{e} Paris 6, Facult\'{e} de physique, 4 place Jussieu, 75252 Paris Cedex 05, France}

\ead{renaud.boussarie@th.u-psud.fr, bertrand.b.ducloue@jyu.fi, lech.szymanowski@ncbj.gov.pl and samuel.wallon@th.u-psud.fr}

\begin{abstract}
We study the production of a forward $J/\psi$ meson and a backward jet with a large rapidity separation at the LHC using the BFKL formalism. We compare the predictions given by the Non Relativistic QCD (NRQCD) approach to charmonium prediction and by the Color Evaporation Model. In NRQCD, we find that the $^3 S_1^{\, 8}$ part of the onium wavefunction is completely dominating the process. NRQCD and the color evaporation model give similar results, although a discrepancy seems to appear as the value of the transverse momenta of the charmonium and of the jet decrease.
\end{abstract}

\section{Introduction}

The high energy behaviour of QCD in the perturbative Regge limit is among the important longstanding theoretical questions in particle physics. QCD dynamics in such a limit are usually described using the BFKL formalism~\cite{Fadin:1975cb,Kuraev:1976ge,Kuraev:1977fs,Balitsky:1978ic}, which relies on $k_t$-factorization~\cite{Cheng:1970ef, FL, GFL, Catani:1990xk, Catani:1990eg, Collins:1991ty, Levin:1991ry}. Many processes have been proposed as a way to probe the BFKL resummation effects which result from these dynamics. One of the most promising ones is the production of two forward jets with a large interval of rapidity, as proposed by Mueller and Navelet~\cite{Mueller:1986ey}. Recent $k_t$-factorization studies of Mueller-Navelet  jets~\cite{Colferai:2010wu,Ducloue:2013hia,Ducloue:2013bva,Ducloue:2014koa} were successful in describing such events at the LHC~\cite{CMS-PAS-FSQ-12-002}. \\
We propose to apply a similar formalism to study the production of a forward $J/\psi$ meson and a backward jet with a rapidity gap that is large enough to probe the BFKL dynamics but small enough for the meson to be tagged at LHC experiments such as ATLAS or CMS.
Although $J/\psi$ mesons were first observed more than 40 years ago, the theoretical mechanism for their production is still to be fully understood and the validity of some models remains a subject of discussions (for recent reviews see for example~\cite{Brambilla:2010cs,Bodwin:2013nua}).
In addition, most predictions for charmonium production rely on collinear factorization, in which one considers the interaction of two on-shell partons emitted by the incoming hadrons, to produce a charmonium accompanied by a fixed number of partons. On the contrary, in this work the $J/\psi$ meson and the tagged jet are produced by the interaction of two collinear partons, but with the resummation of any number of accompanying unobserved partons, as usual in the $k_t$-factorization approach. 
\\
Here we will compare two different approaches for the description of charmonium production. First we will use the NRQCD formalism~\cite{Bodwin:1994jh}, in which the charmonium wavefunction is expanded as a series in powers of the relative velocity of its constituents. Next we will apply the Color Evaporation Model (CEM), which relies on the local-duality hypothesis~\cite{Fritzsch:1977ay,Halzen:1977rs}. Finally we will show numerical estimates of the cross section obtained in both approaches. Further details will be provided elsewhere~\cite{us}.

\section{The scattering cross section in $k_t$-factorization}

\begin{figure}[h]
	\centering\hspace{-1cm}\includegraphics[width=14pc]{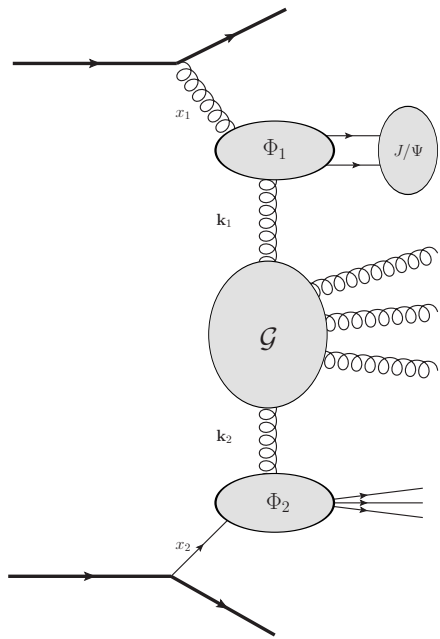}
	\caption{The $k_t$-factorized amplitude for the production of a forward $J/\psi$ meson and a backward jet.}
	\label{Fig:kt}
\end{figure}	
Within the $k_t$-factorization approach for inclusive processes, one writes the cross section as the convolution in transverse momenta of $t$-channel gluons of the impact factor $\Phi_1$ for $J/\psi$ meson production, the impact factor $\Phi_2$ for the production of the backward jet and the BFKL Green's function $\mathcal{G}\,,$ as illustrated in Fig.~\ref{Fig:kt}. Each impact factor contains itself the convolution, in the longitudinal momentum fraction of a parton from the incoming hadron, of a parton distribution function (PDF) with the vertex for the fusion of this parton and a $t$-channel gluon into a $J/\psi$ or a jet. Depending on the quantum numbers of the $c \bar{c}$ pair from which the charmonium will be produced, the upper impact factor may take into account the production of a real gluon. In that case since this gluon will not be tagged, its contribution will be integrated out.

Thus, introducing the azimuthal angles $(\phi_{J/\psi},\phi_{\rm jet})$, the rapidities $(y_{J/_\psi},y_{\rm jet})$ and the transverse momenta $(\mathbf{k}_{J/\psi},\mathbf{k}_{\rm jet})$, one can write the differential cross section as follows :

\begin{eqnarray}
\frac{\mathrm{d}\sigma}{\mathrm{d}|\mathbf{k}_{J/\psi}|\mathrm{d}|\mathbf{k}_{\rm jet}|\mathrm{d}y_{J/\psi}\mathrm{d}y_{\rm jet}} = \int \! \mathrm{d}\phi_{J/\psi} \int \! \mathrm{d}\phi_{\rm jet} \int \! \mathrm{d}^2\mathbf{k}_1 \mathrm{d}^2\mathbf{k}_2 \, \mathcal{G} ( \mathbf{k}_1, \, \mathbf{k}_2, \, \hat{s} ) \\ \nonumber \Phi_1 ( \mathbf{k}_{J/\psi}, \, x_{J/\psi}, \, -\mathbf{k}_1 ) \, \, \Phi_2 ( \mathbf{k}_{\rm jet}, \, x_{\rm jet}, \, \mathbf{k}_2 ).	
\end{eqnarray}

\section{Charmonium production in the Non Relativistic QCD formalism}

The NRQCD formalism is based on the static approximation. Basically, one postulates that the charmonium production can be factorized into two parts. First, the production of an on-shell $c\bar{c}$ pair is computed using the usual Feynman diagram perturbative methods. Then their binding into a charmonium state is encoded in a non-perturbative quarkonium wave-function. Said wavefunction is expanded in terms of the relative velocity $v \sim \frac{1}{\mathrm{log}M}$ of the quarkonium's constituents. In the case of an $S$-state charmonium $J/\psi$ with zero orbital momentum one expands it as follows :
\begin{equation}
\left| \Psi \right\rangle = O(1) \left| Q \bar{Q} \left[^3S_1^{(1)} \right]  \right\rangle + O(v) \left| Q \bar{Q} \left[^3S_1^{(8)} \right] g \right\rangle + O(v^2).
\end{equation}
The first term in this expansion corresponds to the production of a quarkonium from a $c \bar{c}$ pair in a color singlet $S^{(1)}$ state. Due to charge parity conservation, the emission of an additional gluon must be taken into account in the hard part. However, in the second term this additional gluon is included in the wavefunction so it does not appear in the hard part which will then contain only the production of a $c \bar{c}$ pair in a color octet $S^{(8)}$ state.
In the inclusive process studied here and to the first order in $v$, both contributions should be included in the cross section.

\subsection{The color singlet contribution}

\begin{figure}[h]
\centering\includegraphics[width=14cm]{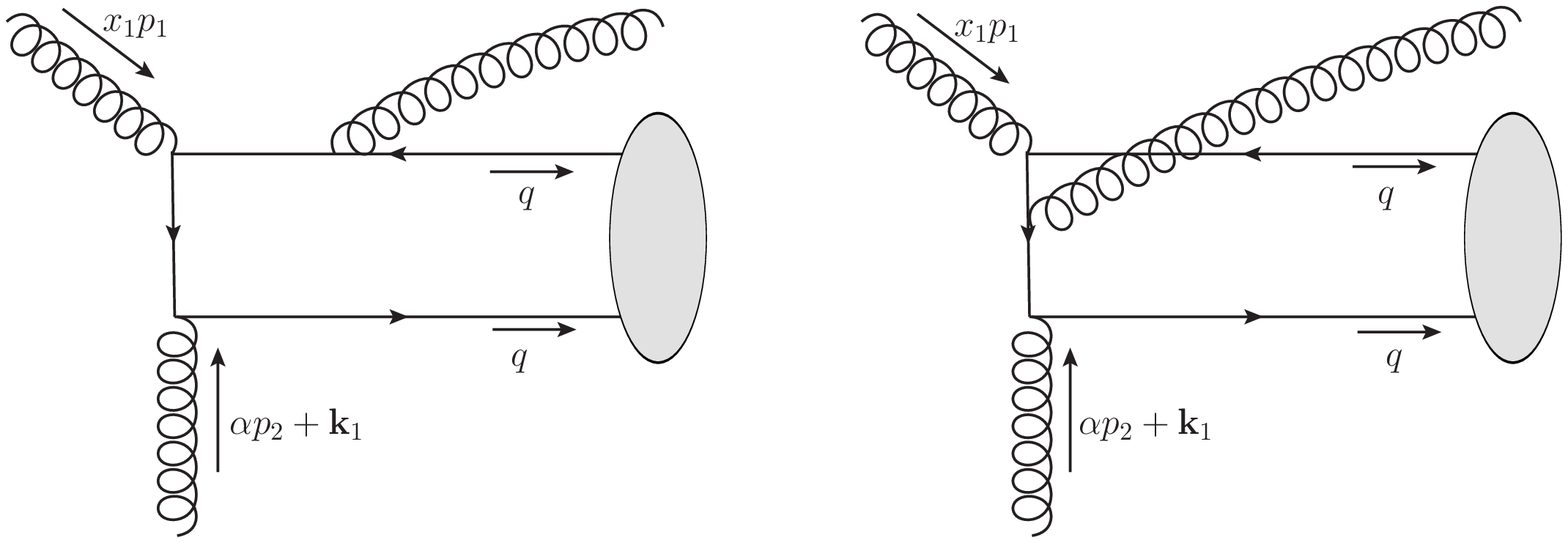}
\caption{Two examples out of the six diagrams contributing to $J/\psi$ production from a $c\bar{c}$ pair in the color singlet state.}
\label{Fig:CSM}
\end{figure}

In this case, the hard part consists of six Feynman diagrams, of which two are illustrated in Fig.~\ref{Fig:CSM}, computed using the color singlet $c \bar{c}$ to $J/\psi$ transition vertex obtained from the NRQCD expansion 
\begin{equation}
v_{\alpha}^i(q_2)\, \bar{u}_{\beta}^j(q_1) \rightarrow \frac{\delta^{ij}}{4N_c} \left( \frac{\langle \mathcal{O}_1 \rangle_{J/\psi}}{m} \right)^{\frac{1}{2}} \left[ \hat{\varepsilon}_{J/\psi}^* \left( \hat{k}_{J/\psi} +M \right) \right]_{\alpha, \, \beta}\,.
\end{equation}
In this equation, $i$ and $j$ are color indices, $\alpha$ and $\beta$ are spinor indices, while $\varepsilon_{J/\psi}$ and $k_{J/\psi}$ are respectively the $J/\psi$ polarization vector and momentum. The $\frac{1}{4N_c}$ factor comes from the projection on spinor indices and on the color singlet. We denote as $m$  the charm quark mass and $M$  the mass of the meson. In the lowest orders in NRQCD one can assume that $M=2m$. One also assumes that the quark and the antiquark carry the same momentum $q$, so that $k_{J/\psi}=2q$, with $q^2=m^2$.
The operator $\mathcal{O}_1$ arises from the non relativistic hamiltonian, and its vacuum expectation value can be fixed by a fit to data. Indeed, it appears for example in the $J/\psi \rightarrow \mu^+ \mu^-$ decay rate.

\subsection{The color octet contribution}
\begin{figure}[h]
\centering\includegraphics[width=14cm]{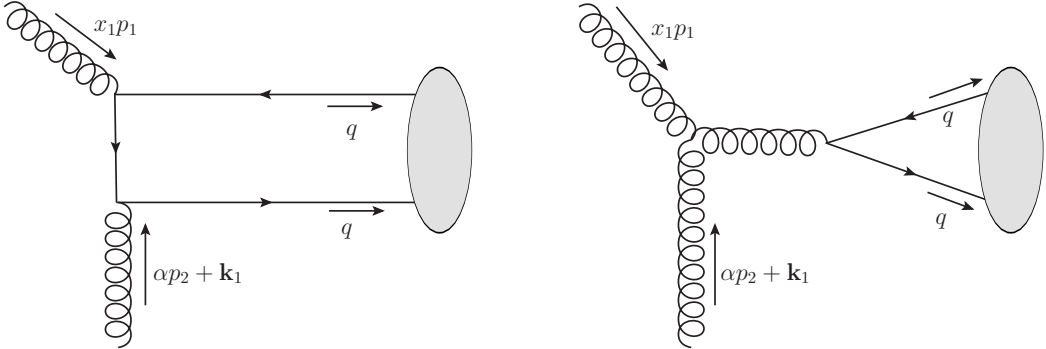}
\caption{Two examples out of the three diagrams contributing to $J/\psi$ production from a $c\bar{c}$ pair in the color octet state.}
\label{Fig:COM}
\end{figure}

The computation of the hard part in the color octet case~\cite{Cho:1995vh,Cho:1995ce} is done in a similar way. It consists of three Feynman diagrams, with two examples shown in Fig.~\ref{Fig:COM}. We use the color octet $c \bar{c}$ to $J/\psi$ transition vertex 
\begin{equation}
\left[ v^i_{\alpha}(q_2) \, \bar{u}^j_{\alpha}(q_1) \right]^a \rightarrow \frac{t^a_{ij}}{4N_c} \left( \frac{\langle \mathcal{O}_8 \rangle_{J/\psi}}{m} \right)^{\frac{1}{2}} \left[ \hat{\varepsilon}_{J/\psi}^* \left( \hat{k}_{J/\psi} +M \right) \right]_{\alpha, \, \beta}\,,
\end{equation}
where the vacuum expectation value of $\mathcal{O}_8$ needs to be determined using experimental data.

\section{The color evaporation model}
\begin{figure}[h]
\psfrag{AA}{\raisebox{-.48cm}{\rotatebox{90}{$\underbrace{\rule{1.cm}{0pt}}$}}}

\psfrag{AAA}{\hspace{-.2cm}\raisebox{-.4cm}{\rotatebox{90}{$\underbrace{\rule{1.cm}{0pt}}$}}}	\centering\includegraphics[width=14cm]{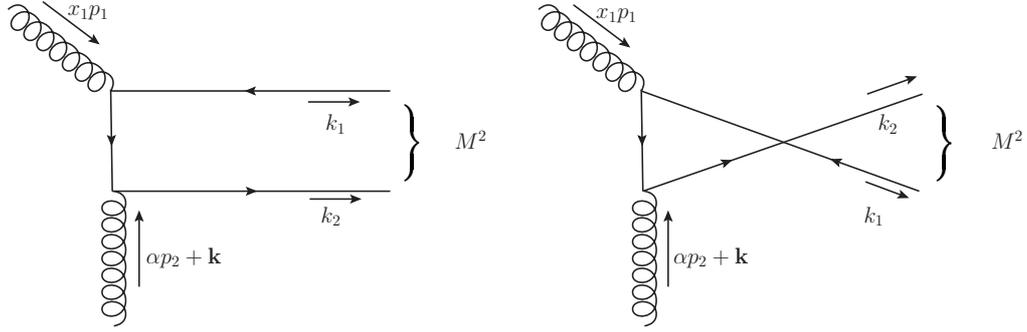}
	\caption{Two examples out of the three diagrams contributing to $J/\psi$ production in the color evaporation model.}
	\label{Fig:CEM}
\end{figure}

While the NRQCD formalism relies on a postulated factorization, the CEM relies on the so-called local duality hypothesis. One assumes that a heavy quark pair $Q \bar{Q}\,,$ with an invariant mass below twice the one of the lightest meson that contains a single heavy quark, will produce a bound $Q \bar{Q}$ state in $\frac{1}{9}$ of the cases, independently of its color. The $\frac{1}{9} = \frac{1}{1+ \left( N_c^2-1 \right) }$ factor accounts for the probability for the quark pair to eventually form a colorless state after a series of randomized soft interactions between its production and its confinement. In the case of a charm quark, the upper limit for the invariant mass corresponds to the threshold $2\,m_D$ for the production of a pair of $D$ mesons. 
The resulting bound state will correspond to any possible heavy quarkonium. One assumes that the repartition between them is universal. \\
In other words the cross section for the production of a $J/\psi$ meson will be a fraction $F_{J/\psi}$ of the cross section for the production of a $c \bar{c}$ pair with an invariant mass $M$ between $2m_c$ and $2m_D$, summed over spins and colors 
\begin{equation}
\sigma_{J/\psi} = F_{J/\psi} \int_{4m_c^{\, 2}}^{4m_D^{\, 2}} dM^2 \frac{d\sigma_{c\bar{c}}}{dM^2}\,.
\end{equation}
Where $F_{J/\psi}$ is supposed to be process-independent and needs to be fitted to data. The diagrams to be computed are similar to the color octet case. Let us however emphasize that the quark and the antiquark no longer carry the same momentum, as required to cover the whole range in allowed invariant masses. This is illustrated in Fig.~\ref{Fig:CEM}.

\section{Numerical results}

\begin{figure}[h]
\psfrag{CS}{\scalebox{0.6}{Color singlet}}
\psfrag{CO}{\scalebox{0.6}{Color octet}}
\psfrag{CEM}{\scalebox{0.6}{Color evaporation model}}
\psfrag{sigma}{\raisebox{0.2cm}{\scalebox{0.7}{$ \displaystyle\frac{\mathrm{d}\sigma}{\mathrm{d}|\mathbf{k}_{J/\psi}| \mathrm{d}|\mathbf{k}_{\rm jet}|\mathrm{d}Y} \left[ \mathrm{nb.GeV^{-2}} \right]$}}}
\psfrag{Y}{\footnotesize$Y$}
\begin{minipage}{7cm}
	\includegraphics[width=7cm]{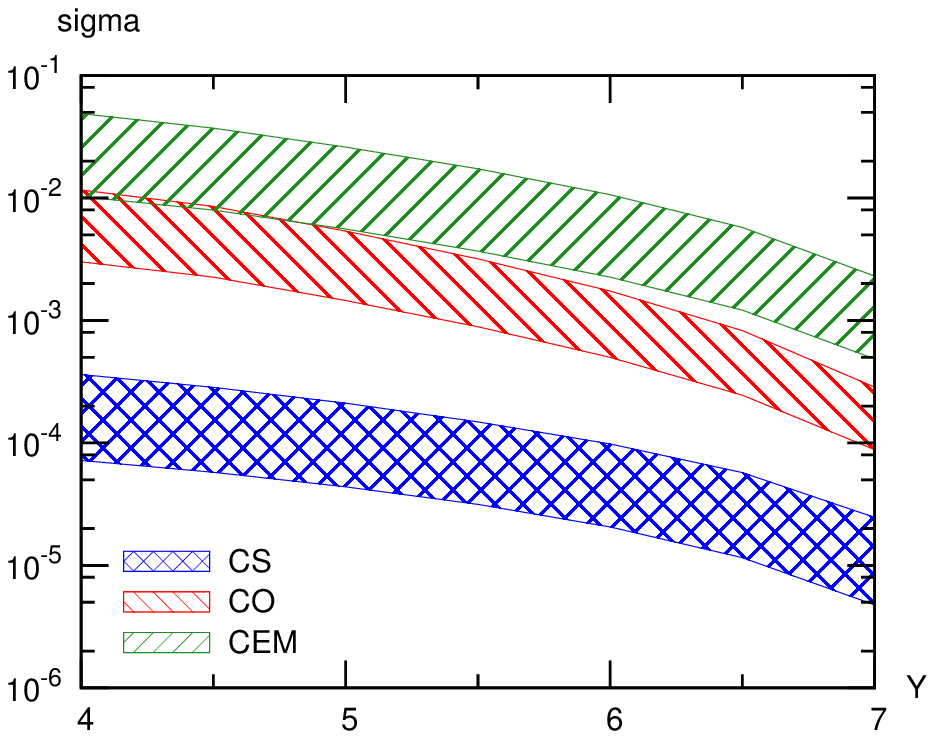}
	\centering\footnotesize $|\mathbf{k}_{J/\psi}|=|\mathbf{k}_{\rm{jet}}|=10$ GeV
\end{minipage}
\hspace{1cm}
\begin{minipage}{7cm}
	\includegraphics[width=7cm]{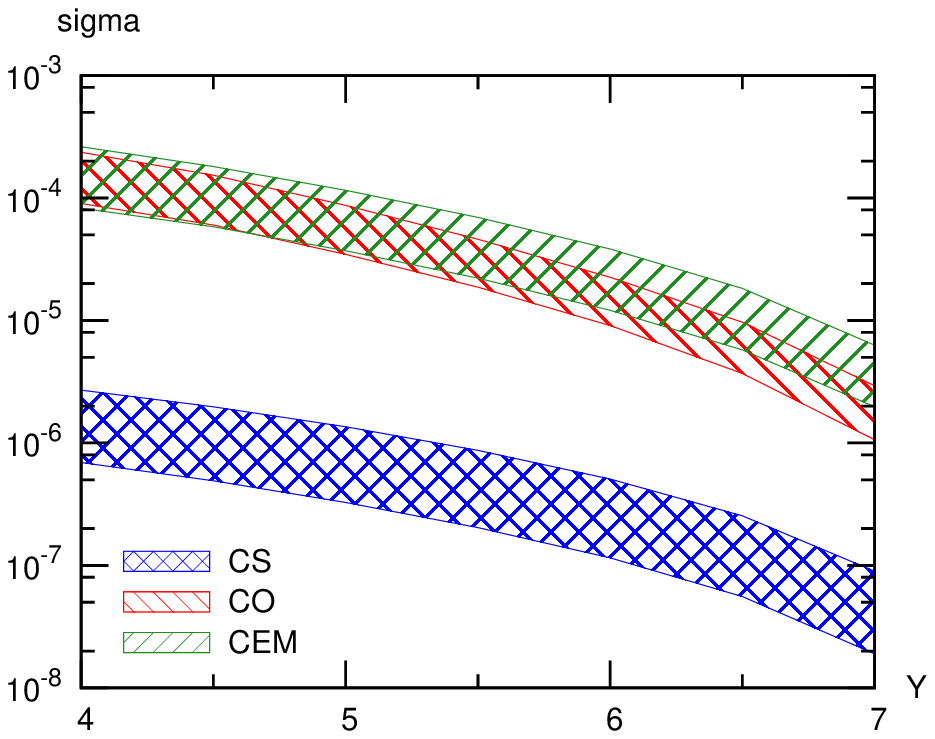}
	\centering\footnotesize $|\mathbf{k}_{J/\psi}|=|\mathbf{k}_{\rm{jet}}|=20$ GeV
\end{minipage}

\vspace{0.5cm}
\begin{minipage}{7cm}
	\includegraphics[width=7cm]{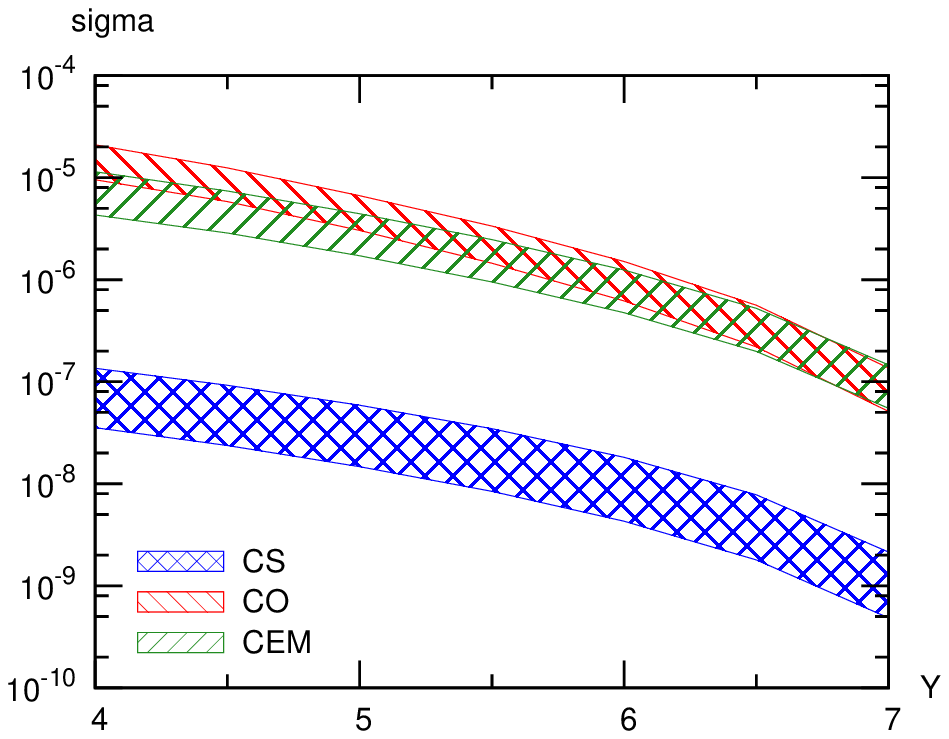}
	\centering\footnotesize $|\mathbf{k}_{J/\psi}|=|\mathbf{k}_{\rm{jet}}|=30$ GeV
\end{minipage}
\label{Fig:Num}
\caption{Differential cross section as a function of $Y$ obtained in NRQCD and in the color evaporation model for three values of $p_T\equiv|\mathbf{k}_{J/\psi}|=|\mathbf{k}_{\rm{jet}}|$.}
\end{figure}

We can now combine the leading order charmonium production vertex obtained above with the BFKL Green's function and the jet vertex. Our implementation is very similar to Ref.~\cite{Ducloue:2013bva}, in particular we use the next-to-leading order jet vertex and the BFKL Green's function at next-to-leading logarithmic accuracy and we use the same scale setting. We note that to perform a complete next-to-leading order study of this process, one would need to compute the NLO corrections to the charmonium production vertex, which could be sizable.
In Fig.~\ref{Fig:Num} we show our results for the cross section as a function of the rapidity separation between the jet and the $J/\psi$, $Y \equiv y_{J/\psi}-y_{\rm jet}$. We use the rapidity cuts $0<y_{J/\psi}<2.5$ and $-4.5<y_{\rm jet}<0$, which are similar to the acceptances for $J/\psi$ and jet tagging at ATLAS and CMS for example.
Here we fix $|\mathbf{k}_{J/\psi}|=|\mathbf{k}_{\rm{jet}}|\equiv p_T$ and we show results for $p_T=10$, 20 and 30 GeV.
For the NRQCD calculation we use the same values for $\langle \mathcal{O}_1 \rangle$ and $\langle \mathcal{O}_8 \rangle$ as in Ref.~\cite{Hagler:2000eu}, where they were determined by comparing a $k_t$-factorization calculation with experimental data. The value of the CEM parameter $F_{J/\psi}$ extracted from data depends on several details of the calculation, such as the PDF parametrization used. In Ref.~\cite{Bedjidian:2004gd}, values between 0.0144 and 0.0248 are quoted. Here we use a value of 0.02 which is approximately in the center of this interval.
We observe from Fig.~\ref{Fig:Num} that in the NRQCD formalism the color singlet contribution is almost negligible compared to the color octet contribution. The cross section in the color evaporation model is of the same order of magnitude as in the NRQCD case, but the two calculations seem to have different behaviours with the kinematics: the decrease of the cross section with increasing $Y$ is slightly more pronounced in the NRQCD approach, while the CEM calculation shows a stronger variation with $p_T$.

\ack
We thank E.~M.~Baldin, A.~V.~Grabovsky and J.-P.~Lansberg for discussions.
B. Duclou\'{e} acknowledges support from the Academy of Finland, Project No. 273464. This work was partially supported by the PEPS-PTI PHENODIFF, the PRC0731 DIFF-QCD, the Polish Grant NCN No. DEC-2011/01/B/ST2/03915, the ANR PARTONS (ANR-12-MONU-0008-01),  the COPIN-IN2P3 Agreement and the Theorie-LHC France intiative.
This work was done using computing resources from CSC -- IT Center for Science in Espoo, Finland.

\end{document}